\documentclass[twocolumn,showpacs,preprintnumbers,amsmath,amssymb,showkeys,prd]{revtex4}

\usepackage{graphicx}
\usepackage{dcolumn}
\usepackage{bm}

\begin{document}

\preprint{APS/123-QED}

\title{Masses and Mixing of $cq \bar{q} \bar{q}$ Tetraquarks \\
Using Glozman-Riska Hyperfine Interaction}

\author{V. Borka Jovanovi\' c}
\email{vborka@vin.bg.ac.yu}
\affiliation{Laboratory of Physics (010), Vin\v ca Institute of Nuclear Sciences, P.O. Box 522\\
11001 Belgrade, Serbia}

\date{\today}

\begin{abstract}
In this paper we perform a detailed study of the masses and mixing
of the single charmed scalar tetraquarks: $cq \bar{q} \bar{q}$. We
also give a systematic analysis of these tetraquark states by weight
diagrams, quantum numbers and flavor wave functions. Tetraquark
masses are calculated using four different fits. The following
SU(3)$_\mathrm{F}$ representations are discussed:
$\overline{15}_\mathrm{S}$, $\bar{3}_\mathrm{S}$, $6_\mathrm{A}$ and
$\bar{3}_\mathrm{A}$. We use the flavor-spin Glozman-Riska
interaction Hamiltonian with SU(3) flavor symmetry breaking. There
are 27 different tetraquarks composed of a charm quark $c$ and of
the three light flavors $u, d, s$: 11 cryptoexotic (3
D$_\mathrm{s}^{+}$, 4 D$^{+}$, 4 D$^{0}$) and 16 explicit exotic
states. We discuss D$_\mathrm{s}$ and its isospin partners in the
same multiplet, as well as all the other four-quark states. Some
explicit exotic states appear in the spectrum with the same masses
as D$_\mathrm{s}^{+}$(2632) in $\overline{15}_\mathrm{S}$ and with
the same masses as D$_\mathrm{s}^{+}$(2317) in $6_\mathrm{A}$
representation, which confirm the tetraquark nature of these states.
\end{abstract}



\maketitle

\section{Introduction}

The meson D$_\mathrm{s}^{+}$(2317), discovered in 2003. year in high
energy electron-positron collisions at SLAC (Stanford Linear
Accelerator Center) by the BABAR group \cite{aube03} and confirmed
by BELLE experiments \cite{mika04}, possesses a mass of 2317 MeV,
some 170 MeV lighter than expected, at least according to prevalent
theories of quark interactions. Hence physicists need a new
explanation of how a charm quark attached to an antistrange quark
should have this particular mass. In general, D$_\mathrm{s}$ and D
mesons are a class of particles, each consisting of a charm quark
attached to a light antiquark. The BABAR detection group at SLAC
\cite{aube03} responsible for the experimental discovery suggests
that the D$_\mathrm{s}^{+}$(2317) might be a novel particle made of
four quarks. Meson D$^{0}$(2308) was discovered by BELLE group
\cite{abe04}. The mass difference between strange
D$_\mathrm{s}^{+}$(2317) and nonstrange D$^{0}$(2308) meson (9 MeV)
is at least ten times below the expected value of $m_\mathrm{s}$ -
$m_\mathrm{u}$ mass difference. Also experimentally state
D$_\mathrm{s}^{+}$(2632) (discovered in SELEX experiments
\cite{evdo04}) does not fit into former theoretical predictions
because it is too light to be an (radial) excitation of the
D$_\mathrm{s}^{+}$(2317).

Jaffe \cite{jaff77a} suggested the possible existence of
four-quark states for light flavor dimesons and made predictions
for tetraquark spectroscopy. In Ref. \onlinecite{jaff77b} it is
also provided a framework for a quark-model classification of the
many two-quark-two-antiquark states.

In Refs. \onlinecite{beve03a,beve03b,beve06} the
D$_\mathrm{s}^{+}$(2317) meson is explained as a scalar $c \bar{s}$
system: van Beveren \emph{et al.} claim that in their model,
assuming that the meson is indeed a charm-antistrange combination,
the mass comes out in the right range if the strong-nuclear-force
interactions responsible for the creation and annihilation of extra
quark-antiquark pairs are taken into account. According to van
Beveren and Rupp \cite{beve04} and Barnes \emph{et al.}
\cite{barn04}, the D$_\mathrm{s}^{+}$(2632) resonance, being 0.52
GeV heavier than the D$_\mathrm{s}$ ground state, could turn out to
be the first radial excitation of the D$_\mathrm{s}$(2112) meson. On
the other hand, Terasaki and Hayashigaki
\cite{tera04b,tera06,haya05} have assigned the
D$_\mathrm{s}^{+}$(2317) to the T$_{3}$ = 0 member of the isotriplet
which belong to the lighter class of four-quark $cq \bar{q} \bar{q}$
mesons and have investigated the decay rates of the members of the
same multiplet. Also in Refs. \onlinecite{tera04a,tera05,haya04} it
is shown that it might be expected that the measured
D$_\mathrm{s}^{+}$(2317) is an exotic state with the structures of a
four-quark. Liu \emph{et al.} \cite{liu04} argue that the
D$_\mathrm{s}^{+}$(2632) resonance may be a member of a scalar
tetraquark multiplet. The possible tetraquark nature of the three
mentioned mesons is discussed e.g. in Refs.
\onlinecite{nico04,niel06,dmit05,dmit06a,dmit06b}.

The three charmed scalar mesons: D$_\mathrm{s}^{+}$(2317),
D$^{0}$(2308) and D$_\mathrm{s}^{+}$(2632) does not fit well into
predictions of the quark model because of these three reasons:
\newline
(i) absolute mass of the D$_\mathrm{s}^{+}$(2317) is 170 MeV below
mass predicted from the quark model for the scalar $c \bar{s}$
meson,
\newline
(ii) the small mass gap between D$_\mathrm{s}^{+}$(2317) and
D$^{0}$(2308) is puzzling and leads to a new model for these states.
\newline
(iii) the state D$_\mathrm{s}^{+}$(2632) does not fit into former
theoretical predictions because it is too light to be an (radial)
excitation of the D$_\mathrm{s}^{+}$(2317).

These three dissimilarities influenced giving some theoretical
proposals about the possible structure of the mesons
D$_\mathrm{s}^{+}$(2317), D$^{0}$(2308) and
D$_\mathrm{s}^{+}$(2632). According to this, we analyze the
possibility that these three states (or some of them) are
tetraquarks.

In this work we perform a schematic study of the mass splitting of
the single charmed $cq \bar{q} \bar{q}$ tetraquarks in the SU(3)
flavor representations. In Section II we construct the wave
functions of mentioned tetraquarks. Then we present the flavor-spin
Glozman-Riska interaction Hamiltonian. The formalism of calculating
SU(3) flavor symmetry breaking corrections to the flavor-spin
interaction energy is presented in Section III. Also it discusses
meson and baryon fit and numerical analysis. The light and heavy
meson and baryon experimental masses are fitted with aim to
calculate the constituent quark masses and then to calculate
tetraquark masses from our theoretical model. We discuss masses with
Glozman-Riska (GR) \cite{gloz96} hyperfine interaction (HFI).
Equations that correspond to our theoretically predicted masses are
given for all 27 $cq \bar{q} \bar{q}$ states, as well as their
numerical values. The quark model of confinement cannot reproduce
the spin-dependent hyperfine splitting in the hadron spectra without
additional contributions from a hyperfine interaction. That is why
we take into account GR hyperfine interaction. We include mass
mixing effects for particles with the same quantum numbers and show
it in mass spectra. The last section is a short summary.

\section{Analysis and Method}

Tetraquarks with charm quantum number C = 1 and with three light
flavors are grouped by the same properties, into multiplets with the
same baryon number, spin and intrinsic parity. If a particle belongs
to a given multiplet, all of its isospin partners (the same isotopic
spin magnitude T and different 3-components T$_{3}$) belong to the
same multiplet.

\begin{figure*}
\centering
\includegraphics[width=0.8\textwidth]{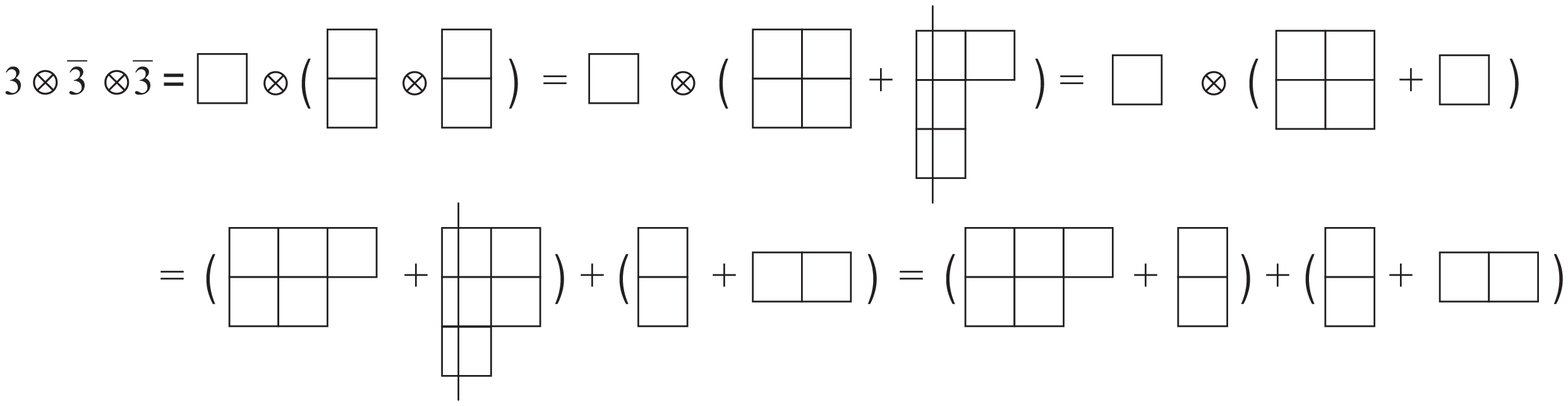}
\caption{Young diagrams for SU(3)$_\mathrm{F}$ multiplets
according to $3\otimes \bar{3} \otimes \bar{3}$ =
($\overline{15}_\mathrm{S}$ + $\bar{3}_\mathrm{S})$ +
($\bar{3}_\mathrm{A}$ + $6_\mathrm{A}$). Tetraquarks with quark
content $cq \bar{q} \bar{q}$ form four multiplets: two
anti-triplets, one anti-15-plet and one sextet.}
\label{fig01}
\end{figure*}

The flavor SU(3) decomposition of the 27 possible $cq \bar{q}
\bar{q}$ combinations is given in Figure \ref{fig01} by the
Young diagrams. Under the transformation of SU(3)$_\mathrm{F}$,
the charm quark is singlet. The numbers 15, 3, 3 and 6 are
dimensions of Young diagrams and they designate the number of
particles in the same group i.e. SU(3) flavor multiplet.
Particles belonging to the same multiplet have the same
baryonic number, spin and intrinsic parity. They also have
similar masses.

\begin{figure*}
\centering
\includegraphics[width=0.7\textwidth]{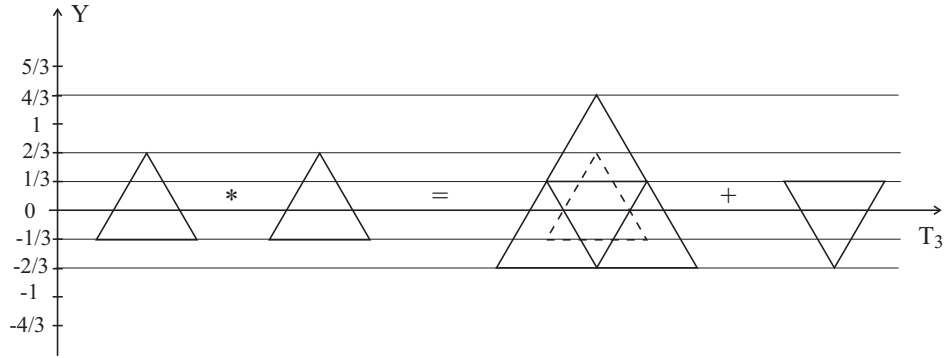}
\caption{Weight diagrams for the product $\bar{3} \otimes
\bar{3} = \bar{6} + 3$. The ordinate shows hypercharge Y and
abscissa 3-component T$_\mathrm{3}$ of isotopic spin
magnitude.}
\label{fig02}
\end{figure*}

\begin{figure*}
\centering
\includegraphics[width=0.7\textwidth]{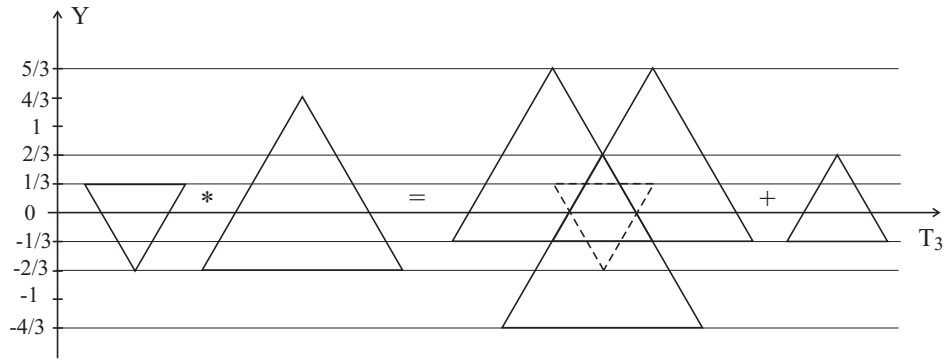}
\caption{The same as in Figure \ref{fig02}, but for the product
$3 \otimes \bar{6} = \overline{15}_\mathrm{S} +
\bar{3}_\mathrm{S}$.}
\label{fig03}
\end{figure*}

\begin{figure*}
\centering
\includegraphics[width=0.7\textwidth]{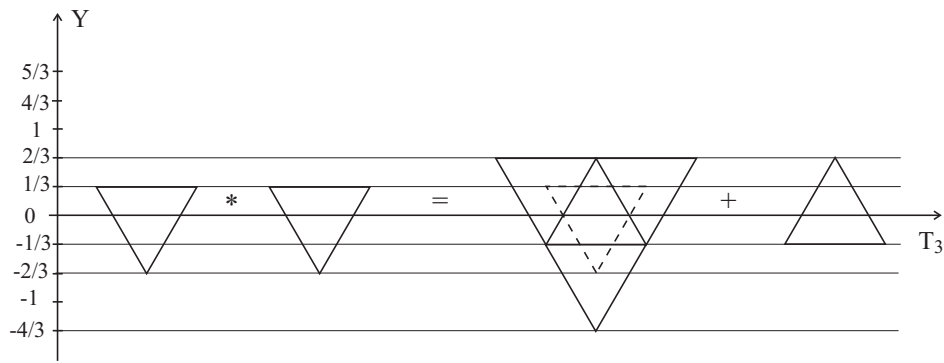}
\caption{The same as in Figure \ref{fig02}, but for the product
$3 \otimes 3 = 6_\mathrm{A} + \bar{3}_\mathrm{A}$.}
\label{fig04}
\end{figure*}

Weight diagrams which represent the following product: $3
\otimes \left( {\bar{3} \otimes \bar{3}} \right) = 3 \otimes
\left( {\bar{6} + 3} \right) = \left( {3 \otimes \bar{6}}
\right) + \left( {3 \otimes 3} \right) = \left(
{\overline{15}_\mathrm{S} + \bar {3}_\mathrm{S}} \right) +
\left( {6_\mathrm{A} + \bar{3}_\mathrm{A}} \right)$ are given
in Figures $\ref{fig02} - \ref{fig04}$.

In these weight diagrams ordinate shows hypercharge Y:

\begin{equation}
Y = B + S + C,
\label{equ01}
\end{equation}

\noindent where B is baryonic number (1/3 for quark, -1/3 for
antiquark), S is strangeness (-1 for $s$ quark, 1 for $\bar{s}$
quark) and C is charm  (1 for $c$ quark, -1 for $\bar{c}$ quark).
So, for tetraquarks with one $c$ quark attached to one light quark
and two light antiquarks, we have: $B = {\frac{{1}}{{3}}} +
{\frac{{1}}{{3}}} - {\frac{{1}}{{3}}} - {\frac{{1}}{{3}}} = 0$ and C
= 1. Also, for electric charge of a particle, we have:

\begin{equation}
Q = T_\mathrm{3} + {\frac{{1}}{{2}}}Y = T_\mathrm{3} +
{\frac{{1}}{{2}}}\left( {S + 1} \right)
\label{equ02}
\end{equation}

We plot the eigenvalues of T$_\mathrm{3}$ and Y that occur for the
quarks in a representation as points in the T$_\mathrm{3}$ - Y
plane. We first combine two of the antiquarks. The quantum numbers Y
and T$_\mathrm{3}$ are additive and thus their values for a $\bar{q}
\bar{q}$ state are obtained by simply adding the values for
$\bar{q}$ and $\bar{q}$. The points in the weight diagram for the
$\bar{3} \otimes \bar{3}$-representation are thus obtained by taking
every point of one antiquark diagram to be the origin of another
antiquark diagram. Figure $\ref{fig02}$ shows that the nine $\bar{q}
\bar{q}$ combinations arrange themselves into two SU(3) multiplets,
where the 3 is symmetric and the $\bar{6}$ is antisymmetric under
interchange of the two antiquarks. Then we add the third quark
triplet. The final decomposition is displayed in Figures
$\ref{fig03}$ and $\ref{fig04}$. The subscripts S and A on the
multiplets indicate that the flavor states are symmetric
($\overline{15}_\mathrm{S}$ and $\bar{3}_\mathrm{S}$) or
antisymmetric ($6_\mathrm{A}$ and $\bar{3}_\mathrm{A}$) under
interchange of the last two antiquarks.

\begin{figure*}
\centering
\includegraphics[width=0.6\textwidth]{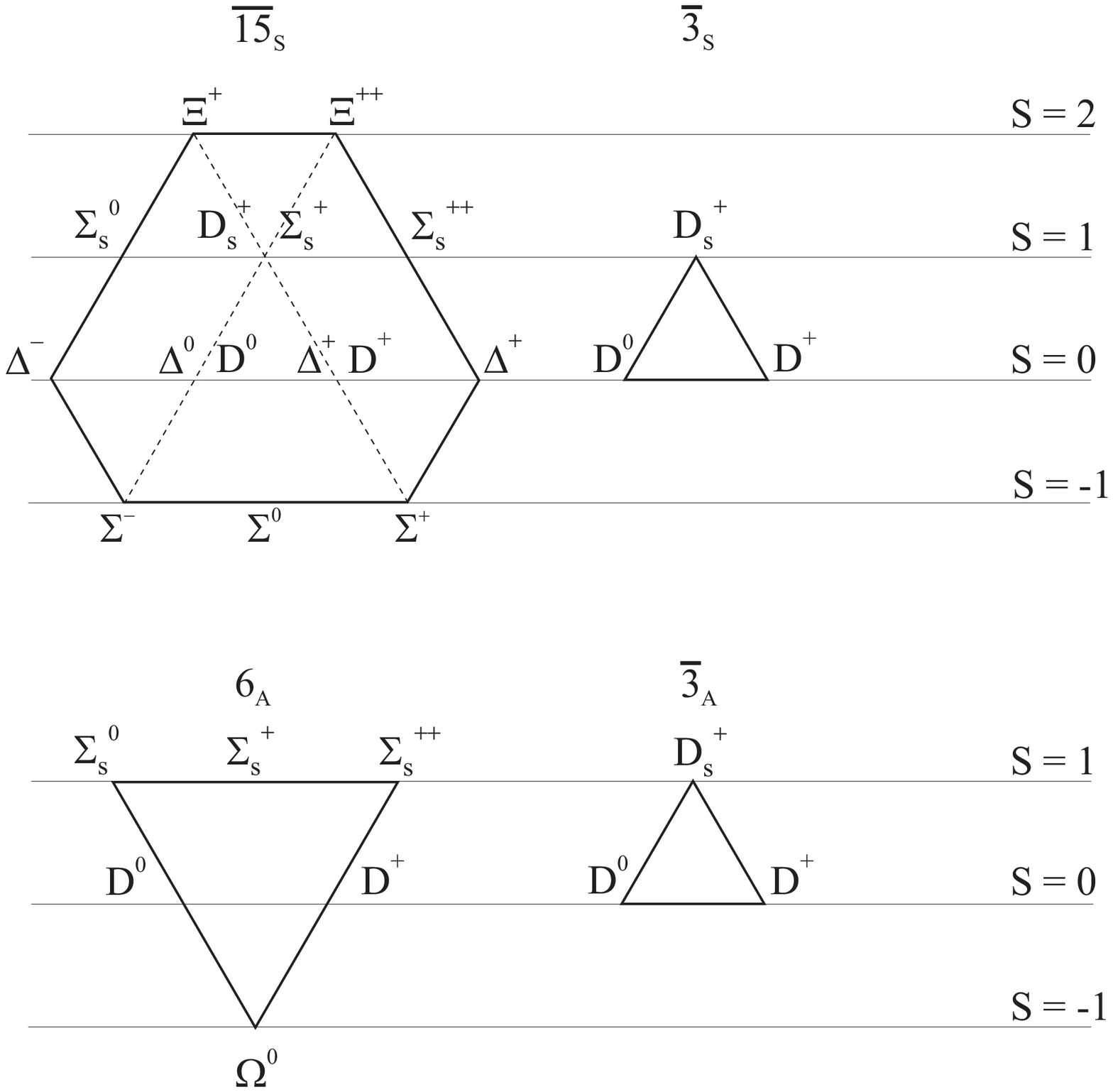}
\caption{Symmetric (above) and antisymmetric (below) tetraquark
multiplets in the SU(3)$_\mathrm{F}$ representation, with a label
for each tetraquark and with given strangeness S (indicated on the
right side).}
\label{fig05}
\end{figure*}

Knowing quantum numbers for the set of 27 scalar tetraquarks,
they are classified in groups as shown in Figure $\ref{fig05}$.
We denote the states with strangeness S = 2 as $\Xi$, with S =
1 as $\Sigma_\mathrm{s}$ (T = 1) and D$_\mathrm{s}$ (T = 0),
with S = 0 as $\Delta$ (T = 3/2) and D (T = 1/2) and with S =
-1 as $\Sigma$.

From the weight diagrams we read off the quark content of the
tetraquarks. The four-quark content, as well as quantum numbers,
calculated for all 27 states in the following SU(3) representations:
$\overline{15}_\mathrm{S}$, $\bar{3}_\mathrm{S}$, $6_\mathrm{A}$ and
$\bar{3}_\mathrm{A}$ are given in Table \ref{tab01}. There is mixing
between states from symmetric multiplets $\overline{15}_\mathrm{S}$
and $\bar{3}_\mathrm{S}$, and also between antisymmetric multiplets
$6_\mathrm{A}$ and $\bar{3}_\mathrm{A}$, while symmetric and
antisymmetric multiplets do not mix with each other. Mixing is due
to the same quantum numbers: electric charge Q($e$), third isospin
projection T$_\mathrm{3}$, isospin T and strangeness S. According to
Table \ref{tab01}, mixed states are D$_\mathrm{s}^{+}$, D$^{+}$ and
D$^{0}$ from symmetric multiplets and  D$^{+}$ and D$^{0}$ from
antisymmetric multiplets. Which mixed state belongs to the
$\overline{15}_\mathrm{S}$ and which to the $\bar{3}_\mathrm{S}$ is
arbitrary at present and, in fact, the physical particle may be some
superposition of the two states. The same applies for mixed states
from $6_\mathrm{A}$ and $\overline{3}_\mathrm{A}$. The flavor wave
functions, requiring orthogonality between each state, are given in
Table \ref{tab02}. It is known that a meson is composed of a quark
and an antiquark, but as can be seen from Table \ref{tab02},
experimentally detected states D$_\mathrm{s}^{+}$(2317) and
D$_\mathrm{s}^{+}$(2632) in addition to $c \bar{s}$ also have $u
\bar{u}$, $d \bar{d}$ and $s \bar{s}$ combinations whose probability
is determined by the square of the coefficient in front of each
combination. In case of D$^{0}$(2308), besides $c \bar{u}$ there are
also $d \bar{d}$ and $s \bar{s}$ combinations. These facts clearly
indicate the tetraquark components in wave functions of the three
mentioned states.

\begin{table*}
\centering
\caption{The four-quark content and quantum numbers of
scalar $cq \bar{q} \bar{q}$ tetraquarks distributed in
SU(3)$_\mathrm{F}$ multiplets.}
\begin{ruledtabular}
\begin{tabular}{lllcccc}
multiplet & tetraquark  & quark & electrical charge & isospin projection & isospin & strangeness \\
& label & content&  Q $(e)$ & T$_\mathrm{3}$& T & S \\
\hline
& & & & & & \\
\raisebox{-0.10ex}[0cm][0cm]{$\overline{15}_\mathrm{S}$}
& $\Xi^{++}$ & $c u \bar{s} \bar{s}$ & 2& 1/2& 1/2& 2 \\
& $\Xi^{+}$ & $c d \bar{s} \bar{s}$ & 1& -1/2& 1/2& 2 \\
& $\Sigma_\mathrm{s}^{++}$ & $c u \bar{d} \bar{s}$ & 2& 1& 1& 1 \\
& $\Sigma_\mathrm{s}^{+}$ & $c q \bar{q} \bar{s}$ & 1& 0& 1& 1 \\
& $\Sigma_\mathrm{s}^{0}$ & $c d \bar{u} \bar{s}$ & 0& -1& 1& 1 \\
& D$_\mathrm{s}^{+} $ & $c q \bar{q} \bar{s}$ & 1& 0& 0& 1 \\
& $\Delta^{++}$ & $c u \bar{d} \bar{d}$ & 2& 3/2& 3/2& 0 \\
& $\Delta^{+}$ & $c q \bar{q} \bar{d}$ & 1& 1/2& 3/2& 0 \\
& $\Delta^{0}$ & $c q \bar{q} \bar{u}$ & 0& -1/2& 3/2& 0 \\
& $\Delta^{-}$ & $c d \bar{u} \bar{u}$ & -1& -3/2& 3/2& 0 \\
& D$^{+}$ & $c q \bar{q} \bar{d}$ & 1& 1/2& 1/2& 0 \\
& D$^{0}$ & $c q \bar{q} \bar{u}$ & 0& -1/2& 1/2& 0 \\
& $\Sigma^{+}$ & $c s \bar{d} \bar{d}$ & 1& 1& 1& -1 \\
& $\Sigma^{0}$ & $c s \bar{u} \bar{d}$ & 0& 0& 1& -1 \\
& $\Sigma^{-}$ & $c s \bar{u} \bar{u}$ & -1& -1& 1& -1 \\
& & & & & & \\
\hline
& & & & & & \\
\raisebox{-0.10ex}[0cm][0cm]{$\bar{3}_\mathrm{S}$}
& D$_\mathrm{s}^{+}$ & $c q \bar{q} \bar{s}$ & 1& 0& 0& 1 \\
& D$^{+}$ & $c q \bar{q} \bar{d}$ & 1& 1/2& 1/2& 0 \\
& D$^{0}$ & $c q \bar{q} \bar{u}$ & 0& -1/2& 1/2& 0 \\
& & & & & & \\
\hline
& & & & & & \\
\raisebox{-0.10ex}[0cm][0cm]{$6_\mathrm{A}$}
& $\Sigma_\mathrm{s}^{++}$ & $c u \bar{d} \bar{s}$ & 2& 1& 1& 1 \\
& $\Sigma_{s}^{+}$ & $c q \bar{q} \bar{s}$ & 1& 0& 1& 1\\
& $\Sigma_\mathrm{s}^{0}$ & $c d \bar{u} \bar{s}$ & 0& -1& 1& 1\\
& D$^{+}$ & $c q \bar{q} \bar{d}$ & 1& 1/2& 1/2& 0 \\
& D$^{0}$ & $c q \bar{q} \bar{u}$ & 0& -1/2& 1/2 & 0 \\
& $\Omega^{0}$ & $c s \bar{u} \bar{d}$ & 0& 0& 0& -1 \\
& & & & & & \\
\hline
& & & & & & \\
\raisebox{-0.10ex}[0cm][0cm]{$\bar{3}_\mathrm{A}$}
& D$_\mathrm{s}^{+}$ & $c q \bar{q} \bar{s}$ & 1& 0& 0& 1 \\
& D$^{+}$ & $c q \bar{q} \bar{d}$ & 1& 1/2& 1/2& 0 \\
& D$^{0}$ & $c q \bar{q} \bar{u}$ & 0& -1/2& 1/2& 0 \\
& & & & & & \\
\end{tabular}
\end{ruledtabular}
\label{tab01}
\end{table*}

\begin{table*}
\centering
\caption{The flavor wave functions of scalar $cq \bar{q}
\bar{q}$ tetraquarks distributed in SU(3)$_\mathrm{F}$ multiplets,
with mixing between states with the same quantum numbers.}
\begin{ruledtabular}
\begin{tabular}{llc}
multiplet & tetraquark & flavor wave function \\
\hline
& & \\
\raisebox{-0.10ex}[0cm][0cm]{$\overline{15}_\mathrm{S}$} & $\Xi^{++}$& $c u \bar{s} \bar{s}$ \\
& $\Xi^{+}$& $c d \bar{s} \bar{s}$ \\
& $\Sigma_\mathrm{s}^{++} $& $- {\frac{{1}}{{\sqrt{2}}} } c u \left( {\bar{d} \bar{s} + \bar{s} \bar{d}} \right)$ \\
& $\Sigma_\mathrm{s}^{+} $& ${\frac{{1}}{{2}}}c\left( {u\left( {\bar
{s}\bar{u} + \bar{u} \bar{s}} \right) - d\left( {\bar{d} \bar{s} + \bar{s} \bar{d}} \right)} \right)$ \\
& $\Sigma_\mathrm{s}^{0}$& ${\frac{{1}}{{\sqrt{2}}} }cd\left( {\bar{s} \bar{u} + \bar{u} \bar{s}} \right)$ \\
& $\Delta^{++}$& $ c u \bar{d} \bar{d} $ \\
& $\Delta^{+}$& $ {\frac{{1}}{{\sqrt{3}}} }c\left( { - u\left(
{\bar{u}\bar{d} + \bar{d} \bar{u}} \right) + d\bar{d} \bar{d}} \right) $ \\
& $\Delta^{0}$& $ {\frac{{1}}{{\sqrt{3}}} }c\left( { - d\left(
{\bar{u} \bar{d} + \bar{d} \bar{u}} \right) + u\bar{u} \bar{u}} \right) $ \\
& $\Delta^{-}$& $ c d \bar{u} \bar{u} $ \\
& $\Sigma^{+}$& $cs \bar{d} \bar{d}$ \\
& $\Sigma_\mathrm{s}^{0}$& $- {\frac{{1}}{{\sqrt{2}}} }cs \left( {\bar{u} \bar{d} + \bar{d} \bar{u}} \right)$ \\
& $\Sigma_\mathrm{s}^{-} $& $c s \bar{u} \bar{u}$ \\
& & \\
\hline
& & \\
\raisebox{-0.10ex}[0cm][0cm] {$\overline{15}_\mathrm{S}$ and
$\bar{3}_\mathrm{S}$ mixed states} &
D$_\mathrm{s}^{+}$($\overline{15}_\mathrm{S}$ --
$\bar{3}_\mathrm{S}$) & $ {\frac{{1}}{{2\sqrt{2}}} }c\left( {u\left(
{\bar{s} \bar{u} + \bar{u} \bar{s}} \right) + d\left( {\bar {d} \bar
{s} + \bar{s} \bar{d}} \right) - 2s \bar{s} \bar{s}} \right) $;
${\frac{{1}}{{2\sqrt{2}}} }c \left( {u \left( {\bar{s} \bar{u} +
\bar{u} \bar{s}} \right) + d \left( {\bar{d} \bar{s} + \bar{s} \bar{d}} \right) + 2s \bar{s} \bar{s}} \right) $ \\
& D$^{+}$($\overline{15}_\mathrm{S}$ -- $\bar{3}_\mathrm{S}$) & $
{\frac{{1}}{{2\sqrt{6}}} }c\left( {u\left( {\bar {u} \bar{d} +
\bar{d} \bar{u}} \right) - 3s\left( {\bar{d}\bar {s} +
\bar{s}\bar{d}} \right) + 2d \bar{d} \bar{d}} \right) $; $
{\frac{{1}}{{2\sqrt{2}}} }c \left( {u \left( {\bar{u} \bar{d} +
\bar{d} \bar{u}} \right) + s \left( {\bar{d} \bar{s} + \bar{s} \bar{d}} \right) + 2d \bar{d} \bar{d}} \right) $\\
& D$^{0}$($\overline{15}_\mathrm{S}$ -- $\bar{3}_\mathrm{S}$) &
${\frac{{1}}{{2\sqrt {6}}} }c\left( { - d\left( {\bar{u} \bar{d} +
\bar{d} \bar{u}} \right) + 3s\left( {\bar{s} \bar{u} + \bar{u}
\bar{s}} \right) - 2u \bar{u} \bar{u}} \right)$; $
{\frac{{1}}{{2\sqrt{2}}} }c \left( {d\left( {\bar{u} \bar{d} +
\bar{d} \bar{u}} \right) + s \left( {\bar{s} \bar{u} + \bar{u} \bar{s}} \right) + 2u \bar{u} \bar{u}} \right) $ \\
& & \\
\hline
& & \\
\raisebox{-0.10ex}[0cm][0cm]{$6_\mathrm{A}$}
& $\Sigma_{s}^{++} $& $ {\frac{{1}}{{\sqrt{2}}} }cu\left( {\bar{d} \bar{s} - \bar{s} \bar{d}} \right) $ \\
& $\Sigma_{s}^{+} $& $ {\frac{{1}}{{2}}}c\left( {u\left( {\bar{s}
\bar{u} - \bar{u} \bar{s}} \right) + d\left( {\bar{d} \bar{s} - \bar{s} \bar{d}} \right)} \right) $ \\
& $\Sigma_{s}^{0}$& $ {\frac{{1}}{{\sqrt{2}}} }cd\left({\bar{s} \bar{u} - \bar{u} \bar{s}} \right) $ \\
& $\Omega^{0}$& $ {\frac{{1}}{{\sqrt{2}}} }cs\left( {\bar{u} \bar{d} - \bar{d} \bar{u}} \right) $ \\
& & \\
\hline
& & \\
\raisebox{-0.10ex}[0cm][0cm]{$\bar{3}_\mathrm{A}$} &
D$_\mathrm{s}^{+}$ & $ {\frac{{1}}{{2}}}c\left( {d\left( {\bar{d}
\bar{s} - \bar{s} \bar{d}} \right) - u\left( {\bar{s} \bar{u} - \bar{u} \bar{s}} \right)} \right) $ \\
& & \\
\hline
& & \\
\raisebox{-0.10ex}[0cm][0cm]{$6_\mathrm{A}$ and $\bar{3}_\mathrm{A}$
mixed states} &
D$^{+}$($6_\mathrm{A}$ -- $\bar{3}_\mathrm{A}$) & $
{\frac{{1}}{{2}}}c\left( {u\left( {\bar{u} \bar{d} - \bar{d}
\bar{u}} \right) + s\left( {\bar{d} \bar{s} -\bar{s} \bar{d}}
\right)} \right) $; $ {\frac{{1}}{{2}}}c\left( {u\left( {\bar{u}
\bar{d} - \bar{d} \bar{u}} \right) - s\left( {\bar{d} \bar{s} - \bar{s} \bar{d}} \right)} \right) $ \\
& D$^{0}$($6_\mathrm{A}$ -- $\bar{3}_\mathrm{A}$) & $
{\frac{{1}}{{2}}}c\left( {d\left( {\bar{u} \bar{d} - \bar{d}
\bar{u}} \right) + s\left( {\bar{s} \bar{u} - \bar{u} \bar{s}}
\right)} \right) $; $ {\frac{{1}}{{2}}}c\left( {s\left(
{\bar{s} \bar{u} - \bar{u} \bar{s}} \right) - d\left( {\bar{u} \bar{d} - \bar{d} \bar{u}} \right)} \right) $ \\
& & \\
\end{tabular}
\end{ruledtabular}
\label{tab02}
\end{table*}

The interaction we use is given by the Hamiltonian operator
\cite{gloz96}:

\begin{equation}
\begin{array}{r}
H_\mathrm{GR}  =  - \mathrm{C_\chi}  \sum\limits_{i < j} {\left( { -
1} \right)^{\mathop \alpha \nolimits_{} ij} \left( {\lambda _i ^F
\lambda _j ^F } \right)\left( {\vec \sigma _i \vec \sigma _j}\right)}\\
\left( { - 1} \right)^{\mathop \alpha \nolimits_{} ij}  = \left\{
{\begin{array}{*{20}c} {\begin{array} {*{20}c}{ - 1,} & {q\bar q}  \\
\end{array}}  \\
{\begin{array}{*{20}c} { + 1,} & {qq \vee \bar q\bar q}  \\
\end{array}}  \\
\end{array}} \right\}
\end{array}
\label{equ03}
\end{equation}

\noindent where $\lambda_{i}^{F}$ are Gell-Mann matrices for flavor
SU(3), $\sigma_{i}$ are the Pauli spin matrices and C$_{\chi}$ is a
constant. We employ this schematic flavor-spin interaction between
quarks and antiquarks which leads to Glozman-Riska HFI contribution
to tetraquark masses \cite{gloz96}:

\begin{equation}
m_\mathrm{\nu ,GR}  = \left\langle {\nu  \uparrow } \right| -
\mathrm{C_\chi} \sum\limits_{i < j = 2}^4 {\left( { - 1}
\right)^{\mathop \alpha \nolimits_{} ij} \frac{{\vec \sigma _i \vec
\sigma _j }}{{m_i m_j }}\left( {\lambda _i ^F \lambda _j ^F }
\right)} \left| {\nu \uparrow } \right\rangle ,
\label{equ04}
\end{equation}
where m$_{i}$ are the constituent quark effective masses:
$m_\mathrm{u}$ = $m_\mathrm{d}$  $\neq$ $m_\mathrm{s}$ and $\nu$ -
flavor wave functions. With the addition of tetraquark masses
$m_\mathrm{\nu,0}$ without influence of Glozman-Riska HFI, finally
for tetraquark masses $m_\mathrm{\nu}$ we have:

\begin{equation}
m_\mathrm{\nu}  = m_\mathrm{\nu,0} + m_\mathrm{\nu ,GR}.
\label{equ05}
\end{equation}

\begin{table*}
\centering
\caption{Masses of scalar $cq \bar{q} \bar{q}$
tetraquarks distributed in SU(3)$_\mathrm{F}$ multiplets, with
mixing between states with the same quantum numbers.
$m_\mathrm{\nu,0}$ are tetraquark masses without influence of GR HFI
and $m_\mathrm{\nu,GR}$ are GR HFI contributions to tetraquark
masses.}
\begin{ruledtabular}
\begin{tabular}{llcc}
multiplet & tetraquark & $m_\mathrm{\nu,0}$ ($m_\mathrm{u}$ = $m_\mathrm{d})$&
$m_\mathrm{\nu,GR}$ ($m_\mathrm{u}$ = $m_\mathrm{d})$ \\
\hline
& & & \\
\raisebox{-0.10ex}[0cm][0cm]{$\overline{15}_\mathrm{S}$} & $\Xi$ &
$m_\mathrm{u}$ + 2$m_\mathrm{s}$ + $m_\mathrm{c}$ & $ -
{\frac{{4}}{{3}}}\mathrm{C_\chi}
\left({{\frac{{2}}{{m_\mathrm{s}^{2}}}} + {\frac{{1}}{{m_\mathrm{u} m_\mathrm{s}}} }} \right)$ \\
& $\Sigma_\mathrm{s}$ & 2$m_\mathrm{u}$+$m_\mathrm{s}$ + $m_\mathrm{c}$ &
$ -{\frac{{4}}{{3}}}\mathrm{C_\chi} \left({{\frac{{1}}{{m_\mathrm{u}^{2}}}} +
{\frac{{2}}{{m_\mathrm{u} m_\mathrm{s}}} }} \right)$ \\
& $\Delta$ & 3$m_\mathrm{u}$ + $m_\mathrm{c}$& $ -4\mathrm{C_\chi} {\frac{{1}}{{m_\mathrm{u}^{2}}}}$ \\
& $\Sigma$ & 2$m_\mathrm{u}$ + $m_\mathrm{s}$ + $m_\mathrm{c}$&
$ - {\frac{{4}}{{3}}}\mathrm{C_\chi}  \left( {{\frac{{1}}{{m_\mathrm{u}^{2}}}} +
{\frac{{2}}{{m_\mathrm{u} m_\mathrm{s}}} }} \right)$ \\
& & & \\
\hline
& & & \\
\raisebox{-0.10ex}[0cm][0cm]{$\overline{15}_\mathrm{S}$ and
$\bar{3}_\mathrm{S}$ mixed states} &
D$_\mathrm{s}$($\overline{15}_\mathrm{S}$ -- $\bar{3}_\mathrm{S}$) &
2$m_\mathrm{u}$ + $m_\mathrm{s}$ + $m_\mathrm{c}$; 3$m_\mathrm{s}$ +
$m_\mathrm{c}$ & $ - {\frac{{4}}{{3}}}\mathrm{C_\chi} \left(
{{\frac{{1}}{{m_\mathrm{u}^{2}}}} + {\frac{{2}}{{m_\mathrm{u}
m_\mathrm{s}}} }} \right) $; $ 4\mathrm{C_\chi} \left(
{{\frac{{1}}{{m_\mathrm{u}^{2}}}} +
{\frac{{1}}{{m_\mathrm{s}^{2}}}} + {\frac{{1}}{{m_\mathrm{u} m_\mathrm{s}}} }} \right) $ \\
& D($\overline{15}_\mathrm{S}$ -- $\bar{3}_\mathrm{S}$) &
3$m_\mathrm{u}$ + $m_\mathrm{c}$; $m_\mathrm{u}$ + 2$m_\mathrm{s}$ +
$m_\mathrm{c}$ & $ -4\mathrm{C_\chi}
{\frac{{1}}{{m_\mathrm{u}^{2}}}}$; ${\frac{{\mathrm{C_\chi}} }
{{6}}} \left({{\frac{{69}}{{m_\mathrm{u}^{2}}}} +
{\frac{{4}}{{m_\mathrm{s}^{2}}}} - {\frac{{1}}{{m_\mathrm{u} m_\mathrm{s}}} }} \right) $ \\
& & & \\
\hline
& & & \\
\raisebox{-0.10ex}[0cm][0cm]{$6_\mathrm{A}$} & $\Sigma_\mathrm{s}$&
2$m_\mathrm{u}$ + $m_\mathrm{s}$ + $m_\mathrm{c}$& $- 8\mathrm{C_\chi} {\frac{{1}}{{m_\mathrm{u} m_\mathrm{s}}} }$ \\
& $\Omega$ & 2$m_\mathrm{u}$ + $m_\mathrm{s}$ + $m_\mathrm{c}$& $- 8\mathrm{C_\chi} {\frac{{1}}{{m_\mathrm{u}^{2}}}}$ \\
& & & \\
\hline
& & & \\
\raisebox{-0.10ex}[0cm][0cm]{$\bar{3}_\mathrm{A}$} & D$_\mathrm{s}$&
2$m_\mathrm{u}$ + $m_\mathrm{s}$ + $m_\mathrm{c}$& $ - 8\mathrm{C_\chi} {\frac{{1}}{{m_\mathrm{u} m_\mathrm{s}}} }$ \\
& & & \\
\hline
& & & \\
\raisebox{-0.10ex}[0cm][0cm]{$6_\mathrm{A}$ and $\bar{3}_\mathrm{A}$
mixed states} &
D($6_\mathrm{A}$ -- $\bar{3}_\mathrm{A}$) & 3$m_\mathrm{u}$ +
$m_\mathrm{c}$; $m_\mathrm{u}$ + 2$m_\mathrm{s}$ + $m_\mathrm{c}$& $- 8\mathrm{C_\chi}
{\frac{{1}}{{m_\mathrm{u}^{2}}}}$; $ - 8\mathrm{C_\chi} {\frac{{1}}{{m_\mathrm{u} m_\mathrm{s}}} }$ \\
& & & \\
\end{tabular}
\end{ruledtabular}
\label{tab03}
\end{table*}

\section{Results}

Using the obtained flavor wave functions of scalar $cq \bar{q}
\bar{q}$ tetraquarks (see Table \ref{tab02}), the tetraquark
masses $m_\mathrm{\nu,0}$ without influence of GR HFI are
determined. GR HFI contributions to the tetraquark masses
$m_\mathrm{\nu,GR}$ are calculated according to the relation
(\ref{equ04}) and the total tetraquark masses by relation
(\ref{equ05}). The corresponding results are given in Table
\ref{tab03}.

The $\chi^{2}$ fit of hadron masses is used to determine masses of
constituent quarks. We performed mass fit for: light mesons ($\pi$,
$K$, $\eta$, $\eta'$, $\rho$, $K^{*}$, $\omega$, $\varphi$), heavy
mesons (D$^{+}$, D$^{0}$, D$^{*0}$, D$^{*+}$, D$_\mathrm{s}^{+}$,
D$_\mathrm{s}^{*+}$, $\eta_\mathrm{c}$, J/$\psi$), light baryons (N,
$\Sigma$, $\Xi$, $\Lambda$, $\Delta$, $\Sigma^{*}$, $\Xi^{*}$,
$\Omega$) and heavy baryons ($\Sigma_\mathrm{c}$,
$\Xi_\mathrm{c}^{+}$, $\Xi_\mathrm{c}^{0}$, $\Lambda_\mathrm{c}$,
$\Sigma_\mathrm{c}^{*}$, $\Omega_\mathrm{c}$). Applying the
Hamiltonian (\ref{equ03}) to the constituent quarks, we obtained the
theoretical meson and baryon masses with the GR contribution
included. Consequently, we have the set of equations
(\ref{equ06})--(\ref{equ13}) for theoretical masses of light
pseudoscalar mesons, light vector mesons, charmed mesons, strange
charmed mesons, double charmed mesons, light baryons - octet, light
baryons - decuplet and heavy baryons, respectively. The
corresponding experimental masses, taken from "Particle Data Group"
site: http://pdg.lbl.gov \cite{PDG06}, are appended to the right
side of each equation:

\begin{equation}
\begin{array}{l}
\left( {m_\pi = } \right)2m_u  - 2C_\chi  \frac{1}{{m_u ^2 }} = 140\ {\rm{MeV}}\\
\left( {m_K  = } \right)m_u  + m_s  - 2C_\chi  \frac{1}{{m_u m_s }} = 494\ {\rm{MeV}}\\
\left( {m_\eta   = } \right)2m_u  - 2C_\chi  \frac{1}{{m_u ^2 }} = 548\ {\rm{MeV}}\\
\left( {m_{\eta '}  = } \right)2m_s  + 16C_\chi  \frac{1}{{m_s ^2 }} = 958\ {\rm{MeV}}
\label{equ06}
\end{array}
\end{equation}

\begin{equation}
\begin{array}{l}
\left( {m_\rho   = } \right)2m_u  + 2C_\chi  \frac{1}{{3m_u ^2 }} = 776\ {\rm{MeV}}\\
\left( {m_{K*}  = } \right)m_u  + m_s  + 2C_\chi  \frac{1}{{3m_u m_s }} = 892\ {\rm{MeV}}\\
\left( {m_\omega   = } \right)2m_u  + 2C_\chi  \frac{1}{{3m_u ^2 }} = 783\ {\rm{MeV}}\\
\left( {m_\varphi   = } \right)2m_s  - 16C_\chi  \frac{1}{{3m_s^2 }} = 1020\ {\rm{MeV}}
\label{equ07}
\end{array}
\end{equation}

\begin{equation}
\begin{array}{l}
\left( {m_{D, \pm }  = } \right)m_u  + m_c  - 2C_\chi  \frac{1}{{m_u m_c }} = 1869\ {\rm{MeV}}\\
\left( {m_{D,0}  = } \right)m_u  + m_c  - 2C_\chi  \frac{1}{{m_u m_c }} = 1865\ {\rm{MeV}}\\
\left( {m_{D*, \pm }  = } \right)m_u  + m_c  + 2C_\chi  \frac{1}{{3m_u m_c }} = 2010\ {\rm{MeV}}\\
\left( {m_{D*,0}  = } \right)m_u  + m_c  + 2C_\chi  \frac{1}{{3m_u m_c }} = 2007\ {\rm{MeV}}
\label{equ08}
\end{array}
\end{equation}

\begin{equation}
\begin{array}{l}
\left( {m_{Ds, \pm } = } \right)m_s  + m_c  + 2C_\chi  \frac{1}{{3m_s m_c }} = 1968\ {\rm{MeV}}\\
\left( {m_{Ds*, \pm } = } \right)m_s  + m_c  + 2C_\chi
\frac{1}{{3m_s m_c }} = 2112\ {\rm{MeV}}
\label{equ09}
\end{array}
\end{equation}

\begin{equation}
\begin{array}{l}
\left( {m_{\eta c} = } \right)2m_c  - 2C_\chi  \frac{1}{{m_c ^2 }} = 2980\ {\rm{MeV}}\\
\left( {m_{J/\psi } = } \right)2m_c  + 2C_\chi  \frac{1}{{3m_c ^2 }}
= 3097\ {\rm{MeV}}
\label{equ10}
\end{array}
\end{equation}

\begin{equation}
\begin{array}{l}
\left( {m_N  = } \right)3m_u  - 8C_\chi  \frac{1}{{m_u ^2 }} = 940\ {\rm{MeV}}\\
\left( {m_\Sigma = } \right)2m_u  + m_s  - C_\chi  \frac{1}{{m_u ^2 }}\left( {1 + 7\frac{{m_u }}{{m_s }}} \right) = 1190\ {\rm{MeV}}\\
\left( {m_\Xi = } \right)m_u  + 2m_s  - C_\chi  \frac{1}{{m_s ^2 }}\left( {1 + 7\frac{{m_s }}{{m_u }}} \right) = 1315\ {\rm{MeV}}\\
\left( {m_\Lambda = } \right)2m_u  + m_s  - C_\chi  \frac{1}{{3 m_u
^2 }}\left( {13 + 11\frac{{m_u }}{{m_s }}} \right) = 1116\
{\rm{MeV}}
\label{equ11}
\end{array}
\end{equation}

\begin{equation}
\begin{array}{l}
\left( {m_\Delta  = } \right)3m_u  - 4C_\chi  \frac{1}{{m_u ^2 }} = 1232\ {\rm{MeV}}\\
\left( {m_{\Sigma *} = } \right)2m_u  + m_s  - 8C_\chi  \frac{1}{{3m_u ^2 }}\left( {\frac{1}{2} + \frac{{m_u }}{{m_s }}} \right) = 1385\ {\rm{MeV}}\\
\left( {m_{\Xi *} = } \right)m_u  + 2m_s  - 8C_\chi  \frac{1}{{3m_s ^2 }}\left( {\frac{1}{2} + \frac{{m_s }}{{m_u }}} \right) = 1530\ {\rm{MeV}}\\
\left( {m_\Omega  = } \right)3m_s  - 4C_\chi  \frac{1}{{m_s ^2 }} =
1672\ {\rm{MeV}}
\label{equ12}
\end{array}
\end{equation}

\begin{equation}
\begin{array}{l}
\left( {m_{\Sigma c} = } \right)2m_u  + m_c  - C_\chi  \frac{1}{{m_u ^2 }}\left( {1 + 7\frac{{m_u }}{{m_c }}} \right) = 2455\ {\rm{MeV}}\\
\left( {m_{\Xi c, + } = } \right)m_u  + 2m_c  - C_\chi  \frac{1}{{m_c ^2 }}\left( {1 + 7\frac{{m_c }}{{m_u }}} \right) = 2470\ {\rm{MeV}}\\
\left( {m_{\Xi c,0} = } \right)m_u  + 2m_c  - C_\chi  \frac{1}{{m_c ^2 }}\left( {1 + 7\frac{{m_c }}{{m_u }}} \right) = 2475\ {\rm{MeV}}\\
\left( {m_{\Lambda c} = } \right)2m_u  + m_c  - C_\chi \frac{1}{{3 m_u ^2 }}\left( {13 + 11\frac{{m_u }}{{m_c }}} \right) = 2285\ {\rm{MeV}}\\
\left( {m_{\Sigma *c} = } \right)2m_u  + m_c  - 8C_\chi  \frac{1}{{3m_u ^2 }}\left( {\frac{1}{2} + \frac{{m_u }}{{m_c }}} \right) = 2520\ {\rm{MeV}}\\
\left( {m_{\Omega c} = } \right)2m_s  + m_c  - 8C_\chi
\frac{1}{{3m_s ^2 }}\left( {\frac{1}{2} + \frac{{m_s }}{{m_c }}}
\right) = 2698\ {\rm{MeV}}
\label{equ13}
\end{array}
\end{equation}

Masses $m_\mathrm{u,d}$, $m_\mathrm{s}$ and $m_\mathrm{c}$ are the
results of the hadron fit. The constant C$_\mathrm{\chi}$ is set so
that the lightest tetraquark from $\bar{3}_\mathrm{A}$ multiplet has
equal mass as D$_\mathrm{s}^{+}$(2317). We performed these
calculations using theoretical and experimental masses of all
particles listed in equations (\ref{equ06})--(\ref{equ13}), except
for mixed states $\eta - \eta'$ (\ref{equ06}) and $\omega - \varphi$
(\ref{equ13}) because the meson octet and singlet mix and the flavor
functions of mixed states are given only in a first approximation
(see \cite{jaff77a}).

The $\chi^{2}$ values for each set of equations for masses are
evaluated as:

\begin{equation}
\chi ^2  = \sum\limits_{i = 1}^N {\frac{{\left( {T_i  - E_i } \right)^2 }}{{\sigma _i ^2 }}},
\label{equ14}
\end{equation}

\noindent where $T_i$ is the model prediction for the hadron mass,
$E_i$ is the experimental hadron mass and $\sigma_i$ is the
uncertainty of the mass. After the values of parameters
($m_\mathrm{u,d}$, $m_\mathrm{s}$, $m_\mathrm{c}$) were obtained by
fitting meson and baryon experimental masses, they were used for
calculation of the tetraquark masses.

The hadron mass fits resulted in the parameter values given in the
Table \ref{tab04}. From the meson fit, calculated masses for $u$ and
$s$ quarks are smaller ($m_\mathrm{u} \approx$ 311 MeV,
$m_\mathrm{s} \approx$ 487 MeV) than those from the baryon fit
($m_\mathrm{u} \approx$ 388 MeV, $m_\mathrm{s} \approx$ 556 MeV).
Due to smaller value of $\chi^{2}$ the masses obtained from the
meson fit are more reliable. Both fits gave the similar values for
constant C$\chi$ ($\sim 7 \times 10^{6}\ \mathrm{MeV^{3}}$).

With parameters from Table \ref{tab04} we calculated tetraquark
masses. The tetraquark masses calculated from meson fit parameters
are given in Table \ref{tab05}. The results from Table \ref{tab05}
show that the isotriplet from $\overline{15}_\mathrm{S}$ has the
same mass as D$_\mathrm{s}^{+}$(2632) and that the isotriplet from
$6_\mathrm{A}$ has the same mass as D$_\mathrm{s}^{+}$(2317).

\begin{table*}
\centering
\caption{The results for masses (in MeV) of constituent
quarks, obtained from hadron masses by $\chi^{2}$ fit. Constant
C$_\mathrm{\chi}$ (in 10$^{6}$ MeV$^{3}$) is set so that the
lightest tetraquark from the $\bar{3}_\mathrm{A}$ multiplet has
equal mass as D$_\mathrm{s}^{+}$(2317).}
\begin{ruledtabular}
\begin{tabular}{lccccc}
hadrons& $m_\mathrm{u}$ (MeV)& $m_\mathrm{s}$ (MeV)&
$m_\mathrm{c}$(MeV)& $\chi^{2}$& C$_\mathrm{\chi}$ (10$^{6}$ MeV$^{3}$) \\
\hline
& & & & & \\
mesons& 311& 487& 1592& 1.04 $\times 10^{-2}$& 7.30 \\
baryons& 388& 556& 1267& 2.29 $\times 10^{-1}$& 7.60 \\
\end{tabular}
\end{ruledtabular}
\label{tab04}
\end{table*}

\begin{table*}
\centering
\caption{The results for masses (in MeV) of scalar $cq
\bar{q} \bar{q}$ tetraquarks distributed in SU(3)$_\mathrm{F}$
multiplets, with mixing between states with the same quantum
numbers, obtained from meson fit. $m_\mathrm{\nu,0}$ (MeV) are
tetraquark masses without influence of GR HFI, $m_\mathrm{\nu,GR}$
(MeV) are GR HFI contributions to tetraquark masses and
m$_\mathrm{\nu}$ (MeV) are the total tetraquark masses.}
\begin{ruledtabular}
\begin{tabular}{llccc}
multiplet & tetraquark & $m_\mathrm{\nu,0}$ (MeV)& $m_\mathrm{\nu,GR}$ (MeV) & $m_\mathrm{\nu} $ (MeV) \\
\hline
& & & & \\
\raisebox{-0.10ex}[0cm][0cm]{$\overline{15}_\mathrm{S}$}& $\Xi$ & 2877& -146& 2731 \\
& $\Sigma_\mathrm{s}$& 2701& -228& 2473 \\
& $\Delta$ & 2525& -301& 2224 \\
& $\Sigma$ & 2701& -228& 2473 \\
& & & & \\
\hline
& & & & \\
\raisebox{-0.10ex}[0cm][0cm]{$\overline{15}_\mathrm{S}$ and
$\bar{3}_\mathrm{S}$ mixed states} &
D$_\mathrm{s}$($\overline{15}_\mathrm{S}$ -- $\bar{3}_\mathrm{S}$)& 2701; 3053& -228; 615& 2473; 3668 \\
& D($\overline{15}_\mathrm{S}$ -- $\bar{3}_\mathrm{S}$)& 2525; 2877& -301; 877& 2224; 3754 \\
& & & & \\
\hline
& & & & \\
\raisebox{-0.10ex}[0cm][0cm]{$6_\mathrm{A}$}& $\Sigma_\mathrm{s}$ & 2701& -384& 2317 \\
& $\Omega$ & 2701& -601& 2100 \\
& & & & \\
\hline
& & & & \\
$\bar{3}_\mathrm{A}$& D$_\mathrm{s}$& 2701& -384& 2317 \\
& & & & \\
\hline
& & & & \\
$6_\mathrm{A}$ and $\bar{3}_\mathrm{A}$ mixed states &
D($6_\mathrm{A}$ -- $\bar{3}_\mathrm{A}$)& 2525; 2877& -601; -384& 1924; 2493 \\
& & & & \\
\end{tabular}
\end{ruledtabular}
\label{tab05}
\end{table*}

GR contribution is positive or negative due to signs of the $(\vec
\sigma_{i} \vec \sigma_{j})$ and $(\lambda_{i}^{F} \lambda_{j}^{F})$
products. It is negative in $\overline{15}_\mathrm{S}$,
$6_\mathrm{A}$ and $\bar{3}_\mathrm{A}$-plets, and for
$\overline{15}_\mathrm{S}$ -- $\bar{3}_\mathrm{S}$ mixed states one
of the mixed states has negative and the other one has positive GR
contribution. The positive GR contribution for two mixed states (see
Table \ref{tab03}) comes out because of the mixing of the states: it
changes the properties and shifts masses from the theoretical
predictions.

For experimentally detected states (D$_\mathrm{s}^{+}$(2317),
D$^{0}$(2308) and D$_\mathrm{s}^{+}$(2632)), hadron fits resulted in
theoretical masses with relatively significant statistical
uncertainties. These uncertainties are mainly due to inaccuracies in
constitutive quark masses obtained using hadron fits (see Table
\ref{tab04}). For example, experimental masses for D$_\mathrm{s}$
and D$_\mathrm{s}^*$ mesons are 1968 MeV and 2112 MeV, but they have
the same quark content (see eq. (\ref{equ09})) and therefore their
constituent quarks have different theoretical masses. Besides,
D$^{0}$(2308) and D$_\mathrm{s}^{+}$(2632) are mixed states (see
Table \ref{tab02}) and therefore their flavor wave functions are
given only in a first approximation. In spite of the uncertainties,
none of the found states with strangeness equal to zero have mass
around $m$ = 2405 MeV, which agrees with the conclusion obtained in
Refs. \onlinecite{niel06} and \onlinecite{brac05} that the state
D$^{0,+}$(2405) (found by the FOCUS collaboration \cite{link04}) is
not a tetraquark, but a normal $c \bar{q}$ state.

\begin{figure}
\centering
\includegraphics[width=0.45\textwidth]{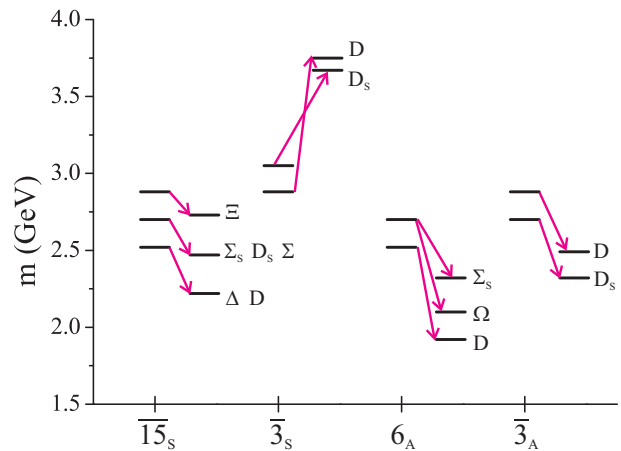}
\caption{Tetraquark mass spectrum from meson fit without (left
column) and with (right column) GR HFI, both with SU(3)$_\mathrm{F}$
symmetry breaking.}
\label{fig06}
\end{figure}

Tetraquark mass spectrum from the meson fit, without and with GR HFI
influence and with SU(3)$_\mathrm{F}$ symmetry breaking is presented
in Figure \ref{fig06}. The general conclusion is that tetraquarks
are arranged in the same way in both spectra: from meson and baryon
fits. The spectra obtained from different fits have a similar
arrangement of particles and if the values of parameters are
changed, the whole spectrum could be shifted towards higher or lower
masses and it could be shrunk or broadened. In both spectra it is
possible to identify D$_\mathrm{s}^{+}$(2317) as the lowest state in
multiplet $\bar{3}_\mathrm{A}$ and D$_\mathrm{s}^{+}$(2632) as a
mixed state from mixing of multiplets $\overline{15}_\mathrm{S}$ and
$\bar{3}_\mathrm{S}$. Also, in both spectra (for example see Table
\ref{fig05}), GR HFI mostly reduces the obtained masses except for
one of D$_\mathrm{s}$ mixed states and one of D mixed states from
$\overline{15}_\mathrm{S}$ -- $\bar{3}_\mathrm{S}$ mixing.

\begin{table*}
\centering
\caption{The results for masses (in MeV) of light scalar
tetraquark nonet, obtained from meson masses by $\chi^{2}$ fit.
Constant C$_\mathrm{\chi}$ = 6.0 $\times 10^{6}$ MeV$^{3}$ is
obtained so the lightest tetraquark $\sigma$ has equal mass as
experimentally state.}
\begin{ruledtabular}
\begin{tabular}{lccccr}
tetraquark& $m_\mathrm{\nu,0}$ ($m_\mathrm{u}$ = $m_\mathrm{d})$&
$m_\mathrm{\nu,GR}$ ($m_\mathrm{u}$ = $m_\mathrm{d})$&
$m_\mathrm{\nu,0}$ (MeV)& $m_\mathrm{\nu,GR}$ (MeV) & $m_\mathrm{\nu} $ (MeV) \\
\hline
& & & & & \\
$\sigma$& $4m_u $& $ - 12C_\chi \frac{1}{{m_u ^2 }}$ & 1244& -744& 500 \\
$f_\mathrm{0}$& $2m_u + 2m_s $ & $ - \frac{2}{3}C_\chi  \left(
{\frac{1}{{m_u ^2 }} + \frac{1}{{m_s ^2 }} + \frac{{16}}{{m_u m_s
}}} \right)$ & 1596& -481& 1115 \\
$\kappa$& $3m_u + m_s $& $ - 6C_\chi  \left( {\frac{1}{{m_u ^2 }} +
\frac{1}{{m_u m_s }}} \right)$& 1420& -610& 810 \\
$a_\mathrm{0}$& $2m_u + 2m_s $& $ - \frac{2}{3}C_\chi  \left(
{\frac{1}{{m_u ^2 }} + \frac{1}{{m_s ^2 }} + \frac{{16}}{{m_u m_s
}}} \right)$& 1596& -481& 1115 \\
\end{tabular}
\end{ruledtabular}
\label{tab06}
\end{table*}

Figure \ref{fig06} may be compared with Figure 3 from Ref.
\onlinecite{dmit06b}, which was plotted for Fermi-Breit (FB)
color-spin HFI from the constituent quark masses obtained in this
way: $m_\mathrm{u}$ was the result of a light meson fit, the mass
difference $m_\mathrm{s}$ - $m_\mathrm{u}$ was taken to be 100 MeV
and $m_\mathrm{c}$ is fitted so that the lightest tetraquark from
the $\bar{3}_\mathrm{A}$ multiplet had equal mass as
D$_\mathrm{s}^{+}$(2317). This comparison shows that obtained
tetraquark masses for both GR and FB HFI are similar, except for
$\overline{15}_\mathrm{S}$ -- $\bar{3}_\mathrm{S}$ mixed states.
Besides, FB HFI is causing additional splitting between
$\Sigma_\mathrm{s}$ (D$_\mathrm{s}$) and $\Sigma$. These
dissimilarities are due to term $(\lambda_{i}^{F} \lambda_{j}^{F})$
in GR and $(\lambda_{i}^{C} \lambda_{j}^{C})$ in FB interaction (F -
flavor, C - color). From this comparison one can see that the forms
of tetraquark spectra with FB and GR interaction are similar, only
they are shifted for some value. This is an important result which
was not expected because FB is a color-spin and GR is a flavor-spin
interaction. The results obtained from these two interactions
confirm that both HFIs give similar results for tetraquark masses.

The lightest $qq \bar{q} \bar{q}$ scalars that are experimentally
known are the $\sigma$ (500), $f_\mathrm{0}$ (980), $\kappa$ (800)
and the $a_\mathrm{0}$ (980). They form an SU(3) flavor nonet.
Already in the seventies Jaffe \cite{jaff77a} suggested the
tetraquark structure of this scalar nonet and proposed a four quark
bag model. Their quark content is given in Refs.
\onlinecite{jaff77a} and \onlinecite{brit05}. It is shown
\cite{brit05} that these mesons fit well in the tetraquark scheme.
We considered them as four quark states and calculated their masses,
which are given in Table \ref{tab06}. As can be seen from Table
\ref{tab06}, GR HFI significantly reduces the theoretical masses of
the light scalar tetraquarks and brings them closer to their
experimental masses. This fact confirmes the conclusion from Ref.
\onlinecite{brit05} about tetraquark nature of these light scalars.

\section{Conclusions}

We have made a systematic analysis of the charm tetraquark
states. Weight diagrams, irreducible representations and flavor
wave functions are shown and analyzed. Detailed method of
calculation is described.

The mass spectrum with mixing of particles with the same quantum
numbers is shown. The discussion how results depend on parameters is
also given.

There are 27 different tetraquarks with C = 1 and with three light
flavors. We calculated mass spectra and wave functions for all 27
states using GR HFI. Among these states there are 11 cryptoexotic (3
D$_\mathrm{s}^{+}$, 4 D$^{+}$, 4 D$^{0}$) and 16 explicit exotic
states. In the tetraquark model it is possible to identify
D$_\mathrm{s}^{+}$(2317) and D$_\mathrm{s}^{+}$(2632) with two
cryptoexotic states in tetraquark spectrum when Glozman-Riska
hyperfine interaction is included. Namely, explicit exotic states,
as for instance isotriplet $\Sigma_\mathrm{s}$
($\Sigma_\mathrm{s}^{++}$, $\Sigma_\mathrm{s}^{+}$,
$\Sigma_\mathrm{s}^{0}$), appear in the spectrum with the same
masses (i.e. at 2317 MeV and at 2632 MeV). One intriguing
possibility is that D$^{0}$(2308) is a tetraquark. Then, there
should be the other tetraquark partner of D$^{0}$(2308) to form an
isospin doublet with isospin T = 1/2 (see Table \ref{tab01}). We
suggest that the recently discovered charm-strange meson
D$^{0}$(2308), with unusual properties, could be a cryptoexotic
tetraquark state $cq \bar{q} \bar{u}$. The tetraquark nature for
scalar charmed mesons D$_\mathrm{s}^{+}$(2317), D$^{0}$(2308) and
D$_\mathrm{s}^{+}$(2632) is confirmed by showing existence of the
tetraquark component in their wave functions.

We also gave estimates for masses of experimentally detected light
scalars $\sigma$ (500), $f_\mathrm{0}$ (980), $\kappa$ (800) and
$a_\mathrm{0}$ (980) and confirmed that they satisfactorily fit in
the tetraquark scheme when GR HFI is included.

If we compare masses with and without hyperfine interactions we can
conclude that mass arrangement of tetraquark flavor multiplets
depends almost entirely on the strong hyperfine interaction. We show
that in both cases of hyperfine interaction (FB and GR) the lowest
lying multiplet is $6_\mathrm{A}$-plet, and the mixing and ordering
of other states is similar in the two models. Maybe, FB and GR are
not the complete effective two-quark interactions, and because of
that theoretical prediction is not the same as the experiment.

We also showed for the first time wave functions and quark content
for all predicted 27 quark states of $cq \bar{q} \bar{q}$
combination. We obtained all masses using GR interaction with two
fits and we showed that GR HFI gave similar results as FB
interaction. More experimental searches for detection of other $cq
\bar{q} \bar{q}$ members especially those exotic ones are needed in
the future.

\begin{acknowledgments}
The author acknowledges support by the Ministry of Science of
Serbia. Also, the author would like to thank V. Dmitra\v{s}inovi\' c
for useful suggestions and support.
\end{acknowledgments}




\begin{thebibliography}{00}   

\bibitem{aube03}
BABAR Collab. (B. Aubert \emph{et al.}), Phys. Rev. Lett.
\textbf{90}, 242001 (2003)

\bibitem{mika04}
BELLE Collab. (Y. Mikami \emph{et al.}), Phys. Rev. Lett.
\textbf{92}, 012002 (2004)

\bibitem{abe04}
BELLE Collab. (K. Abe \emph{et al.}), Phys. Rev. D \textbf{69},
112002 (2004)

\bibitem{evdo04}
SELEX Collab. (A. V. Evdokimov \emph{et al.}), Phys. Rev. Lett.
\textbf{93}, 242001 (2004)

\bibitem{jaff77a}
R. L. Jaffe, Phys. Rev. D \textbf{15}, 267 (1977)

\bibitem{jaff77b}
R. L. Jaffe, Phys. Rev. D \textbf{15}, 281 (1977)

\bibitem{beve03a}
E. van Beveren and G. Rupp, AIP Conf. Proc. \textbf{687}, 86 (2003)

\bibitem{beve03b}
E. van Beveren and G. Rupp, Phys. Rev. Lett. \textbf{91}, 012003
(2003)

\bibitem{beve06}
E. van Beveren, J. E. G. N. Costa, F. Kleefeld and G. Rupp, Phys.
Rev. D \textbf{74}, 037501 (2006)

\bibitem{beve04}
E. van Beveren and G. Rupp, Phys. Rev. Lett. \textbf{93}, 202001
(2004)

\bibitem{barn04}
T. Barnes, F. E. Close, J. J. Dudek, S. Godfrey and E. S. Swanson,
Phys. Lett. B \textbf{600}, 223 (2004)

\bibitem{haya05}
A. Hayashigaki and K. Terasaki, Prog. Theor. Phys. \textbf{114},
1191 (2005)

\bibitem{tera04b}
K. Terasaki, arXiv:hep-ph/0405146 (2004)

\bibitem{tera06}
K. Terasaki, Prog. Theor. Phys. \textbf{116}, 435 (2006)

\bibitem{tera04a}
K. Terasaki, AIP Conf. Proc. \textbf{717}, 556 (2004)

\bibitem{tera05}
K. Terasaki, Prog. Theor. Phys. \textbf{114}, 205 (2005)

\bibitem{haya04}
A. Hayashigaki and K. Terasaki, arXiv:hep-ph/0411285 (2004)

\bibitem{liu04}
Y.-R. Liu, S.-L. Zhu, Y.-B. Dai and C. Liu, Phys. Rev. D
\textbf{70}, 094009 (2004)

\bibitem{nico04}
B. Nicolescu and J. P. B. C. de Melo, arXiv:hep-ph/0407088 (2004)

\bibitem{niel06}
M. Nielsen, R. D. Matheus, F. S. Navarra, M. E. Bracco and A. Lozea,
Nucl. Phys. B, Proc. Suppl. \textbf{161}, 193 (2006)

\bibitem{dmit05}
V. Dmitra\v sinovi\' c, Phys. Rev. Lett. \textbf{94}, 162002 (2005)

\bibitem{dmit06a}
V. Dmitra\v sinovi\' c, Modern Phys. Let. A \textbf{21}, 533 (2006)

\bibitem{dmit06b}
V. Dmitra\v sinovi\' c, Int. J. Mod. Phys. A \textbf{21}, 5625
(2006)

\bibitem{gloz96}
L. Ya. Glozman and D. O. Riska, Phys. Rep. \textbf{268}, 263 (1996)

\bibitem{PDG06}
W.-M. Yao \emph{et al.}, J. Phys. G \textbf{33}, 1 (2006)

\bibitem{brac05}
M. E. Bracco, A. Lozea, R. D. Matheus, F. S. Navarra and M. Nielsen,
Phys. Lett. B \textbf{624}, 217 (2005)

\bibitem{link04}
FOCUS Collab. (J. M. Link \emph{et al.}), Phys. Lett. B
\textbf{586}, 11 (2004)

\bibitem{brit05}
T. V. Brito, F. S. Navarra, M. Nielsen and M. E. Bracco, Phys. Lett.
B \textbf{608}, 69 (2005)


\end{thebibliography}
\end{document}